\documentclass[prc,amsmath,twocolumn,showpacs,superscriptaddress]{revtex4}
\usepackage[dvips]{graphicx}
\usepackage{dcolumn}

\newcommand{\bea}{\begin{eqnarray}}
\newcommand{\eea}{\end{eqnarray}}
\newcommand{\be}{\begin{equation}}
\newcommand{\ee}{\end{equation}}

\renewcommand{\vec}[1]{\boldsymbol{#1}}

\newcommand{\dif}{{\mathrm{d}}}

\newcommand{\dderiv}[1]{\frac{\dif^2}{\dif #1^2}}

\newcommand{\bra}[1]{\langle#1|}
\newcommand{\ket}[1]{|#1\rangle}

\newcommand{\conkvc}{\frac{2\mu_{vc}}{\hbar^2}}
\newcommand{\convc}{\frac{\hbar^2}{2\mu_{vc}}}

\newcommand{\TR}{T_{\vec{R}}}
\newcommand{\Tr}{T_{\vec{r}}}
\newcommand{\coords}{\vec{R},\vec{r},\xi}

\begin{document}

\title{Core Transitions in the Breakup of Exotic Nuclei}

\author{N.~C.~Summers}
 \affiliation{National Superconducting Cyclotron Laboratory,
Michigan State University, East Lansing, Michigan 48824, U.S.A.}
\author{F.~M.~Nunes}
 \email{nunes@nscl.msu.edu}
 \affiliation{National Superconducting Cyclotron Laboratory,
Michigan State University, East Lansing, Michigan 48824, U.S.A.}
 \affiliation{Department of Physics and Astronomy,
Michigan State University, East Lansing, Michigan 48824, U.S.A.}
\author{I.~J.~Thompson}
 \affiliation{Department of Physics, University of Surrey,
 Guildford, GU2 5XH, U.K.}
\date{\today}

\begin{abstract}
An interesting physical process has been unveiled: dynamical
core excitation during a breakup reaction of loosely bound $core+N$ systems.
These reactions are typically used to
extract spectroscopic information and/or astrophysical information.
A new method,   the eXtended Continuum Discretized
Coupled Channel (XCDCC) method, was developed to incorporate, in a consistent way
and to all orders, core excitation in the bound and scattering states
of the projectile,  as well as dynamical excitation of the core
as it interacts with the target. The model predicts cross sections
to specific states of the core. It is applied to
the breakup of $^{11}$Be on $^9$Be at 60 MeV/u, and the calculated
cross sections are in improved agreement with the data.
The distribution of the cross section amongst
the various core states is shown to depend on the reaction model used, and not
simply on the ground state spectroscopic factors.
\end{abstract}


\maketitle

In order to study nuclei at the limit
of stability one needs reliable nuclear reactions models
that incorporate the relevant structure degrees of freedom
in a consistent manner, in particular the continuum.
Theories of nuclear reactions have been repeatedly challenged
with the new avenue of experimental work now possible at
Radioactive Beam Facilities. Amongst the various reactions,
breakup occupies a prominent role. With breakup reactions, one
tries to extract spectroscopic information \cite{fukuda,aumann00} or
capture reaction rates of Astrophysical relevance \cite{trache,ogata,esbensen05}.
In either case, the structure information obtained is model
dependent and assumes a single particle description of the projectile
as a valence nucleon attached to the ground state of the core.
Such a simplification may be the source of lingering disagreements
\cite{summers}.

When charged particle detectors are coupled with gamma arrays,
the states of the core can be disentangled. For the variety of knockout
measurements now available \cite{knockout}, the data are found to contain
both elastic breakup (diffraction) and transfer (stripping) contributions,
which are typically calculated within an eikonal spectator core model
\cite{tostevin99}. Only the nuclear reaction cross sections to particular
states are computed and other effects need to be added {\it a posteriori},
and incoherently (i.e. \cite{aumann00}).
One uses the single particle contributions weighted by the
composition of the initial state predicted by shell model
\cite{tostevin99} and neglects interference between the various
projectile components and dynamical processes with the target.
However, one should expect that, if the energy necessary to excite
the core is small,  there will be dynamical core excitation/de-excitation
during  the reaction with the target.
When a loosely bound projectile cannot be described as a single
particle state, core excited components are mixed in the
projectile scattering states as well as the ground state.
These can interfere during the reaction and modify the single
particle picture.

Exotic  systems of the type $core+N$ where core degrees of freedom may play
a relevant role include $^8$Li, $^8$B, $^{11}$Be, $^{17}$C, $^{17}$F and $^{19}$C.
In addition, one can expect to find in their spectra states
built on excited states of the core. This is the case of a resonance in
$^{11}$Be which is visible at approximately $3.4$ MeV excitation in the spectrum
of \cite{fukuda}, and such states cannot be understood within the single
particle description \cite{capel}. This calls for a formulation of breakup where core
excitation is consistently included.

As a first example, we look at $^9$Be($^{11}$Be,$^{10}$Be)X at 60 MeV/u \cite{aumann00},
a reaction already studied in detail.
The eikonal model predictions for the total cross sections
to particular states are too low:
$\sigma_\mathrm{th}=165$ mb to be compared with $\sigma_\mathrm{exp}=203(31)$ mb
for the ground state of $^{10}$Be, and
$\sigma_\mathrm{th}=9$ mb to be compared with $\sigma_\mathrm{exp}=16(4)$ mb
for the $2^+$ excited state.
Estimates of a Coulomb contribution and inelastic core excitation
presented in \cite{aumann00} provide a possible explanation for the
apparent under-prediction of theory, but the problem has been
awaiting a consistent formulation.
Furthermore, Continuum Discretized Coupled Channel calculations
show that the eikonal approximation does not have the desired level
of accuracy at this energy\cite{tostevin02}. We propose a model where all these
corrections and effects are included in a consistent manner.

A recent preliminary study  \cite{batham} generalizes the spectator core model of
\cite{tostevin99} to include core excitation. Although
the work is performed for nuclear only and within a straight line approximation,
it represents an important improvement over previous efforts because
a core degree of freedom is introduced consistently in the projectile and the
core-target interaction, thus allowing for dynamical core excitation/de-excitation
throughout the reaction.
The initial results in \cite{batham} show an increase of the breakup component of the total cross sections,
but no effect on the stripping part. This is an extremely
useful result, that the stripping component seems to be less affected by
the various mechanisms discussed above.
We will focus here on elastic breakup  (diffraction dissociation) only.

As mentioned above, typically breakup models assume a single particle
description for the projectile.
Only recently are improvements on this approximation being considered.
One impressive improvement consists of describing the projectile
as a three body system \cite{egami04} adequate for nuclei
of Borromean nature such as $^6$He. We pursue an alternative improvement,
which is to  describe the projectile as a multi-component system based
on several core states. In this work, we present the eXtended
Continuum Discretized Coupled Channel (XCDCC) method,
to take into account explicitly core excitation in the
breakup reaction of loosely bound systems.
\begin{figure}[b]
\includegraphics[width=0.25\textwidth]{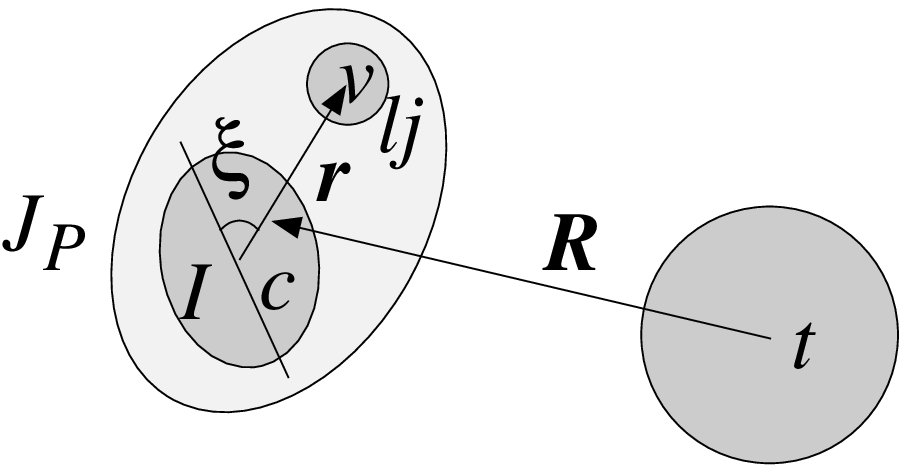}
\caption{\label{coord} Coordinates for the three body breakup reaction. }
\end{figure}

We consider the breakup of a projectile (P), composed of a core (c)
plus a valence particle (v), on a target (T).
The breakup process is described using a three body
Hamiltonian, with core degrees of freedom denoted by $\xi$:
$H^{\mathrm{3b}} = \TR + H_{\mathrm{proj}}(\vec r,\xi) +
U_{vt}(\vec r,\vec R) + U_{ct}(\vec r,\vec R,\xi)$.
The coordinates are illustrated in Fig. \ref{coord}, and, as
Jacobi coordinates, can be used for the full three-body
wavefunction
\bea\label{wf3b}
&&\Psi_{J_TM_T}(\coords) = \nonumber \\ && \sum_{\alpha} \Psi^{J_T}_{\alpha}(R)
  \left[ \left[ Y_L(\hat{\vec{R}}) \otimes \Phi_{J_P}^{in}(\vec{r},\xi) \right]_J \otimes
   \Phi_{J_t} \right]^{J_TM_T}   \!\!\!\! ,
\eea
where $L$ is the projectile-target orbital angular momentum,
$J_t$ the total spin of the target,
$J_T$  the total spin of the three-body system,
and $\alpha = \{L,J_P,J,J_t,i,n\}$.

The projectile states $\Phi_{J_P}(\vec{r},\xi)$ can be either
a bound or a scattering state, with several components. They are
obtained as coupled channels eigenstates of the projectile
Hamiltonian $H_{\mathrm{proj}}=\Tr + V_{vc}(\vec{r}, \xi) + h_{\mathrm{core}}(\xi)$:
\be\label{wfproj}
\Phi_{J_P}^{in}(\vec{r},\xi) = \sum_{a}  u^i_{a:n}(r)
 [[ Y_l(\hat{\vec{r}}) \otimes \chi_s ] _j  \otimes  \varphi_{I}(\xi) ]_{J_P}
 \;,
\ee
expanded in terms of core eigenstates at energies $\varepsilon_{I}$.


A continuum projectile state is characterised by
$\{J_P,i,n\}$, where $i$ refers to the asymptotic energy, and $n$ denotes the channel
with a plane wave component.
It is composed of projectile radial wavefunctions $ f_{a:n}(r;k_{an})$ for each
$a=\{lsj,I\}$ that are solutions of coupled equations which,
since the state of the core can change, are
\bea\label{hccbin}
\left [E_k - \varepsilon_{a} + \varepsilon_{n} +
\convc \left( \dderiv{r} - \frac{l(l+1)}{r^2} \right)\right] f_{a:n}(r;k_{an})  \nonumber \\
= \sum_{a'} V_{a:a'}(r) f_{a':n}(r;k_{a'n}) \;,\;\;\;
\eea
where
the coupling potentials are matrix elements of $V_{vc}(\vec{r},\xi)$.
From these equations we obtain the S-matrix $S_{a:n}$.

Given the importance of continuum-continuum couplings in the CDCC formulation
of breakup for halo nuclei \cite{nunes99}, we need to transform the projectile scattering
states into square integrable functions, otherwise continuum-continuum couplings
would diverge. We define a coupled channel bin as
\be\label{ccbin}
u^i_{a:n}(r) = \sqrt{\frac{2}{\pi (k_{i}{-}k_{i-1})}} \int_{k_{i-1}}^{k_i} \, \dif k ~ e^{-i \delta_n(k)}
f_{a:n}(r;k_{an})
\ee
where $k$ is the core-valence relative momentum,  \linebreak
$k_{an}^2 = k^2{ -} \conkvc [\varepsilon_a{-}\varepsilon_n]$.
From $S_{a:n}$ we obtain  $\delta_n(k)$, the diagonal valence-core phase shift of channel $a=n$.
Coupled channel bins defined in this way are complex.

Substituting the three body wavefunction Eq.~(\ref{wf3b}) into the three body
Schr\"odinger equation \linebreak
$H^{\mathrm{3b}} \Psi_{J_T}= E \Psi_{J_T}$, one obtains an equation set
similar to the standard CDCC equations of \cite{cdcc-theory},
with coupling potentials $U^{J_T}_{\alpha:\alpha'}(R) =
\bra{\alpha} U_{ct}(\vec r,\vec R,\xi)+U_{vt}(\vec r,\vec R) \ket{\alpha'}$
containing both Coulomb and nuclear interactions between the projectile
fragments and the target. The techniques for solving this equation are the same
as in \cite{nunes99}. However there are essentially two differences in these
eXtended CDCC equations: first they span a larger number
of projectile coupled components, corresponding to core excitation, and second
the interaction core-target depends on core degrees of freedom $\xi$.
In our calculations, the couplings $U^{J_T}_{\alpha:\alpha'}(R)$ are further
expanded in multipoles and non-trivial algebra is necessary to simplify
the problem,  but we leave the details on the evaluations of these matrix elements
for a longer publication \cite{summers05}. A new generation of
the code {\sc fresco} \cite{fresco} was developed to incorporate these aspects,
namely coupled channel bins and core excitation CDCC couplings.
The code was further optimized and parallelized so that realistic calculations
could become feasible.

The first application of XCDCC is to the breakup components
of $^9$Be($^{11}$Be,$^{10}$Be)X at 60 MeV/u \cite{aumann00}.
As the spin $s$ of the neutron has an insignificant effect, we set it to $s=0$.
We include only the first excited state of the core as in \cite{nunes96}. For our case
the ground state of $^{11}$Be  contains two components: an s-wave neutron coupled
to $^{10}$Be($0^+$) and a d-wave neutron coupled to $^{10}$Be($2^+$).
The model for n-$^{10}$Be  is based on \cite{nunes96}, but the depths of the
Woods-Saxon interactions need to be adjusted to give
a positive parity bound state energy at $E_{+}=-0.5$ MeV and a
negative parity level at $E_{-}=-0.18$ MeV,  in the $s=0$ approximation.
The needed interaction in the coupled channel model (CC) of
$^{11}$Be is listed in the first row of table \ref{pot}.
It produces a ground state of $^{11}$Be where the neutron is $88.3$\% in the s-wave
and $11.7$\% in the d-wave. Note that these probabilities can be related to the
spectroscopic factors in the shell model which are normalized to particle number \cite{sm}.
For comparison we also want to calculate the breakup cross section using
the previous CDCC approach. In this calculation, each single particle
contribution is multiplied by the corresponding projectile's probability and
all components are added incoherently. Therefore we label this calculations by SPIS
for single particle incoherent sum).  SPIS neglects core excitation during breakup, and uses the single
particle potentials built on the g.s. of the core and its first excited state as
listed in the second and third rows of table \ref{pot}.
These potentials are used to generate the bound states and the whole continuum
(these structure calculations are referred to as SP for single particle).

The neutron-target optical potential parameters are taken from \cite{bg}.
As the target $^9$Be is a spectator, its spin $J_t$  is neglected, as in Refs
\cite{aumann00,knockout,tostevin99,tostevin02,batham}.
For the $^{10}$Be-$^{9}$Be we take the potentials from \cite{jim} that reproduce
unpublished data for the elastic scattering of $^{10}$Be on $^{12}$C at 59 MeV/u. This potential
is directly used in the SP calculations. For the XCDCC calculations, we deformed both
the nuclear and the Coulomb parts using the  deformation lengths consistent
with our structure model for $^{11}$Be \cite{be11nunes} and refit the potential
to reproduce the same elastic distribution. The resulting potential is given in the last row of
table \ref{pot}.

\begin{figure}
\includegraphics[width=0.4\textwidth]{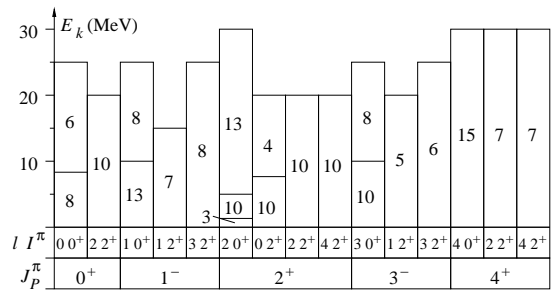}
\caption{\label{modelspace}$^{11}$Be continuum model space.
The number of bins and the energy range are given for each outgoing channel
($l,I^\pi$) for each spin parity combination
of the projectile ($J_P^\pi$).}
\end{figure}

We briefly describe the model space for solving the XCDCC equations.
We take partial waves for $J_P\le4$ organized in
178 bins as schematically shown in Fig. \ref{modelspace}.
The evaluation of the
couplings $U^{J_T}_{\alpha:\alpha'}(R)$ involve an integration in $r$ which is performed
up to $r_{\mathrm{bin}}=70$ fm, and a multipole expansion
which we truncate at $\Lambda_{\mathrm{max}}=4$. The projectile-target relative angular momentum
is taken up to $L_{\mathrm{max}}=3000$ and the corresponding distorted waves
are matched to Coulomb functions at  $R_\mathrm{asym}=500$ fm.
The XCDCC calculations were performed on a SGI Altix 3700.

\begin{table}\begin{tabular}{|c|c||c|c|c|c|c|}\hline
model & interaction                 & V(+)    & $V(-)$       & $R$       & $a$       & $\beta_2$ \\\hline
CC    & $^{10}$Be\{$0^+,2^+$\}+n    & 55.25       & 47.00     & 2.483     & 0.65      & 0.67      \\
SP    & $^{10}$Be\{$0^+$\}+n        & 55.50       & 30.48     & 2.736     & 0.67      & 0.0         \\
SP    & $^{10}$Be\{$2^+$\}+n        & 75.07       & 39.95     & 2.736     & 0.67      & 0.0         \\\hline
model & interaction        & $V$  &  $W$   & $r$ & $a$ & $\beta_2$ \\\hline
CC    & $^{10}$Be-$^{9}$Be & 134.1& 68.15 & 0.75  & 0.8 & 0.67 \\ \hline
\end{tabular}
\caption{\label{pot} Potential parameters for $^{10}$Be+n with and without deformation
of the core. Also given are the parameters for the core-target optical potential.}
\vspace{-0.2cm}
\end{table}

The results are shown in table \ref{results}. Using a single particle model for
$^{11}$Be, and introducing ground state occupation probabilities consistent
with the ones produced
by the CC model, namely $88.3$ \%  s-wave and $11.7$ \%  d-wave,
the predicted cross sections are $109$ mb to the g.s. of the core, and $1$ mb
to the first excited state (model SPIS).  In comparison,
when core excitation is included consistently, the XCDCC calculations
predict exactly the same cross section to the g.s. of the core, but a large
increase (a factor of 8) in the cross section to the $^{10}$Be $2^+$ state.
In both SPIS  and XCDCC breakup, the probabilities for seeing the $2^+$ state,
$P_{2}=\sigma_2/\sigma_\mathrm{tot}$, are much lower than in the ground state.
This reflects the fact that the partial cross
section decreases rapidly with increasing single-particle Q-value for breakup.

The occupation probability (or for that matter the spectroscopic factor) is not an observable,
therefore it is not good practice to compare theory and experiment at this level.
Rather one should include an appropriate reaction model and compare cross sections.
If we take the eikonal prediction for nuclear breakup directly from \cite{aumann00}
and include occupation probabilities consistent with our $^{11}$Be coupled
channel model, we obtain cross sections lower than the XCDCC predictions (see first row
of table \ref{results}). The differences between the theoretical cross sections
predicted in the Eikonal model and the data \cite{aumann00}
were attributed to the Coulomb breakup and inelastic excitation of the core.
\begin{table}\begin{tabular}{|c||c|c||c|}\hline
Model &       $\sigma_{0^+}$& $\sigma_{2^+}$&$\sigma$ \\ \hline
Eikonal &      105  mb         & 3.4 mb     & 108 mb \\
SPIS    &        109  mb        &  1 mb        & 110 mb    \\
XCDCC  &       109 mb         &  8 mb       & 117 mb    \\ \hline
\end{tabular}
\caption{\label{results} $^{11}$Be breakup cross sections for $^{10}$Be$\{0^+,2^+\}$ + n.}
\vspace{-0.2cm}
\end{table}
In XCDCC one can turn off the various couplings to understand their
relative importance. We find that even when core excitation couplings are
not included in the core-target interaction, the cross section to the $2^+$
is increased over the SPIS prediction, due to constructive interference between the projectile's
components in the continuum. There are also nuclear-Coulomb interference effects.
Consequently, inelastic and Coulomb contributions should not be added
incoherently.
These are automatically contained in the XCDCC predictions.

For a meaningful comparison with the data one needs to construct the
stripping component, since the neutron was not detected in the measurement \cite{aumann00}.
However XCDCC produces the elastic breakup component only.
In principle we would like to produce the stripping in the same framework
as the breakup but this is at present not possible.
From the work of Batham {\it et al.} \cite{batham} we learned that the stripping is
hardly affected by core excitation, so we take the stripping contribution
calculated in the eikonal approximation \cite{tostevin05}, based on the same optical
potentials as the XCDCC calculation here presented. The results are summarized
in table \ref{th-exp} and immediately one can see that the
theoretical cross sections agree perfectly with the data.
\begin{table}\begin{tabular}{|c||c|c|c||c|}\hline
 core state    & $\sigma_\mathrm{bu}$ &    $\sigma_\mathrm{st}$ &  $\sigma_\mathrm{th}$  &   $\sigma_\mathrm{exp}$\\ \hline
$0^+$ & 109 mb   &   91 mb     &    200    &    203(31)\\
$2^+$ &   8 mb   &    6 mb     &     14    &     16(4)\\ \hline
\end{tabular}
\caption{\label{th-exp} Comparison of calculated  cross sections,
sum of XCDCC breakup and Eikonal stripping, with the data \cite{aumann00}
for $^9$Be($^{11}$Be,$^{10}$Be)X at 60 MeV/u.}
\vspace{-0.5cm}
\end{table}

Since the core-target optical potential was scaled from that obtained from elastic
scattering of $^{10}$Be on a similar target ($^{12}$C), we test the sensitivity on the
choice of this potential.
For this purpose, we use a microscopically based potential, by folding the NN interaction over
the density of the core, to reproduce the same elastic and inelastic
cross sections for $^{10}$Be as the potential of table \ref{pot}.
The XCDCC breakup cross sections to each core state are not affected, and
neither are the corresponding stripping cross sections.
This demonstrates that details of the core-target optical potential are not important,
as long as similar observables for the core-target interaction are obtained.

Finally, it is necessary to point out that within our model space, there
is no population of the $^{10}$Be $1^-$ and $2^-$ states, in this reaction.
These states are seen in the experiment \cite{aumann00}, but
we expect that these states, which result from neutron excitation from the core, will not interfere
with the results here presented. At present we do not have a $^{10}$Be structure model
that incorporates all $2^+, 1^-, 2^-$ states in a simple way, as these
correspond to breaking the core.

In conclusion, we find that the amount of core excitation is modified in the breakup reaction.
This process occurs both
through constructive interference of various components in the projectile
and through the interaction of the core with the target.
The eXtended Continuum Discretized Coupled Channel method was
developed specifically to handle the problem
of core excitation in the breakup of loosely bound projectiles.
Effects of core excitation
in the projectile bound and scattering states are explicitly taken into account.
Through the interaction with the target, the core can excite or de-excite
during the reaction. We have applied XCDCC to the breakup of $^{11}$Be on
$^9$Be at 60 MeV/u, where the final $^{10}$Be state is identified.
Theoretical predictions within a truncated model space agree with the data for
the cross sections to the first two individual $^{10}$Be states.

As compared to the preliminary calculations  of Batham {\it et al.} \cite{batham},
XCDCC provides more detailed cross sections, namely partial cross sections to each core
state. As a consequence we now understand that the increase of the total cross section
when including core excitation comes mainly from an increase of the core excited component.
Thus one can think of it as production of core excitation.
This process was not understood before.

Despite the computational challenge, XCDCC consists on a significant improvement
to previous theories. Given the possibility of producing core excitation, previous
spectroscopic analyses and extractions of astrophysical S-factors need to be
revisited. Other reactions that can be usefully studied with XCDCC include
$^{11}$Be(p,p$'$)$^{11}$Be*, $^{12}$C($^{11}$Be,$^{10}$Be+n),
the various modes of $^{8}$B breakup, and work along these lines is underway.
So far, the method is limited to the inclusion of
states of the core that can be modelled as collective excitations,
but it could easily be adapted to including a better description of the core, as long as
a complete set of $core+N$ (bound and scattering states) could be obtained.

We thank Jeff Tostevin for providing the stripping cross sections
for comparison with the data and for useful discussions. We also
thank Jim Al-Khalili for providing the data for $^{10}$Be.
This work is supported by NSCL, Michigan State University,
the U.K. EPSRC by grant GR/T28577,
and by the National Science Foundation through grant PHY-0456656.
The calculations presented were performed at the HPCC at
Michigan State University.


\begin{thebibliography}{10}

\bibitem{aumann00}
T. Aumann  {\it et al.}, Phys. Rev. Lett. {\bf 84}, 35 (2000).

\bibitem{fukuda}
N. Fukuda {\it et al.}, Phys.\ Rev.\ C {\bf 70}, 054606 (2004).

\bibitem{trache}
L. Trache {\it et al.}, Phys.\ Rev.\ C {\bf 69}, 032802 (2004).

\bibitem{ogata}
K. Ogata {\it et al.}, arvix.org/nucl-th/0505007, 2005,
submitted to Phys.\ Rev.\ C.

\bibitem{esbensen05}
H. Esbensen, G.~F. Bertsch, and K.~A. Snover,
Phys.\ Rev.\ Lett.\ {\bf 94}, 042502 (2005).

\bibitem{summers}
N.~C. Summers and F.~M. Nunes, J.\ Phys.\ G {\bf 31}, 1437 (2005).

\bibitem{knockout}
P.~G. Hansen and J.~A. Tostevin,
Annu.\ Rev.\ Nucl.\ Part.\ Sci.\ {\bf 53}, 219 (2003).

\bibitem{tostevin99} J.~A. Tostevin, J.\ Phys.\ G {\bf 23}, 735 (1999).

\bibitem{capel}
P. Capel, G. Goldstein, and D. Baye,
Phys.\ Rev.\ C {\bf 70}, 064605 (2004).

\bibitem{tostevin02}
J.~A. Tostevin, Phys.\ Rev.\ C {\bf 66}, 024607 (2002).

\bibitem{batham} P. Batham, I.~J. Thompson, and J.~A. Tostevin,
    Phys.\ Rev.\ C {\bf 71}, 064608 (2005).

\bibitem{egami04}
T. Egami {\it et al.}, Phys.\ Rev.\ C {\bf 70}, 047604 (2004).

\bibitem{nunes99}
F.~M. Nunes and I.~J. Thompson, Phys.\ Rev.\ C {\bf 59}, 2652 (1999).

\bibitem{cdcc-theory}
Y. Sakuragi, M. Yahiro, and M. Kamimura, Prog.\ Theor.\ Phys.\ Suppl. {\bf 89},  136  (1986).

\bibitem{summers05}
N.~C. Summers, F.~M. Nunes, and I.~J. Thompson, {\it in preparation}.

\bibitem{fresco}
I.~J. Thompson, Comput.\ Phys.\ Rep. {\bf 7},  3 (1988).

\bibitem{sm}
H. Esbensen, B.A. Brown and H. Sagawa, Phys. Rev. C {\bf 51}, 1274 (1995).

\bibitem{nunes96}
F.~M. Nunes, J.~A. Christley, I.~J. Thompson, R.~C. Johnson,
and V.~D. Efros, Nucl.\ Phys.\ {\bf A609}, 43 (1996).

\bibitem{bg} F.~D. Becchetti and G.~W. Greenlees,
Phys.\ Rev.\ {\bf 182}, 1190 (1969).

\bibitem{jim}
J.~S. Al-Khalili, J.~A. Tostevin, and J.~M. Brooke,
Phys.\ Rev.\ C {\bf 55}, R1018 (1997).

\bibitem{be11nunes} F.~M. Nunes, I.~J. Thompson, and R.~C. Johnson,
        Nucl.\ Phys.\ {\bf A596}, 171 (1996).

\bibitem{tostevin05}
J.~A. Tostevin, {\it private communication}, MSU 2005.


\end{thebibliography}
\end{document}